\documentclass{elsart3p}

\usepackage{graphicx}

\usepackage{amssymb}

\begin{document}

\begin{frontmatter}

\title{The Directional Dependence of the Lunar Cherenkov Technique for UHE Neutrino Detection}

\author[1]{C.W.~James} \ead{clancy.james@adelaide.edu.au}
\author[1]{R.J.~Protheroe}

\address[1]{
Dept. of Physics, School of Chemistry \& Physics, Univ. of Adelaide, 
SA 5000, AUSTRALIA}

\begin{abstract}
The LUNASKA (Lunar UHE Neutrino Astrophysics with the Square Kilometre Array) project is a
theoretical and experimental project developing the lunar Cherenkov technique for the
next generation of giant radio-telescope arrays. This contribution presents our simulation
results on the directional dependence of the technique for UHE neutrino detection. In
particular, these indicate that both the instantaneous sensitivities and
time-integrated limits from lunar Cherenkov experiments such as those at Parkes,
Goldstone, Kalyazin and ATCA are highly anisotropic. We study the regions of the sky which
have not been probed by either these or other experiments, and present
the expected sky coverage of future experiments with the SKA. Our results show how the
sensitivity of Lunar Cherenkov observations to potential astrophysical sources of
UHE particles may be maximised by choosing appropriate observations dates and antenna-beam
pointing positions.
\end{abstract}

\end{frontmatter}

\section{Introduction}
\label{intro}

Observations of ultra-high energy neutrinos (UHE $\nu$) have been proposed
to resolve the mystery of the origins of the UHE cosmic rays (CR), either
by using the flux (or limits thereon) to discriminate between production models,
or by using observed UHE $\nu$ arrival directions to point back to their source.
Recent results from the Pierre
Auger observatory \cite{AugerScience07} show that the UHE CR flux itself is not
isotropic, indicating that at the highest energies some directional information
is preserved. Thus observations of either UHE $\nu$ or a large increase
in UHE CR statistics could be used to resolve the mystery. For an experiment
aiming to observe either of these particles therefore, the dependence of
experimental sensitivity on UHE particle arrival direction becomes important.

The Lunar Cherenkov technique --- proposed by Dagkesamanskii
and Zheleznykh~\cite{Dagkesamanskii} --- is a method to observe both UHE CR and $\nu$.
By observing the Moon with Earth-based radio-telescopes, the pulses of
microwave-radio radiation produced via the Askaryan effect \cite{askaryan62}
from these particles interacting in the Lunar regolith may be detected.
The first attempt utilised the Parkes radio telescope \cite{Parkes96}, and
subsequent experiments \cite{GLUE,Beresnyak05,Scholten08}
have placed limits on an isotropic flux of UHE neutrinos, though these
are not currently competitive with experiments
such as ANITA \cite{ANITA} and RICE \cite{RICE} except for those of \cite{Scholten08} at the
very highest energies. However,
with the advent of the next generation of giant radio arrays such as LOFAR \cite{LOFAR}
and the Square kilometre Array (SKA), the sensitivity of the technique will increase greatly, and
observations with the SKA are expected to probe the `cosmogenic' neutrino
flux from UHE CR interactions, and could detect a very high rate
of the UHE CR themselves \cite{JP_limits}.

In this contribution we examine the directional-dependence of the sensitivity
of the Lunar Cherenkov technique to UHE $\nu$, and analyse the ability of current experiments (such
as ours with ATCA \cite{LUNASKA_experimental})
to make targeted observations. We calculate a limit on UHE $\nu$ from the Parkes
experiment and GLUE as a function of particle arrival direction, which
combined with approximate dependencies for experiments such as RICE and ANITA,
allows us to identify regions of the primary energy--arrival direction parameter
space which have been relatively unprobed by all current observations. Finally,
we examine the likely sky coverage of future experiments with the ATCA, ASKAP,
and the SKA. Though current simulation methods are appropriate only
to the detection of UHE $\nu$ \cite{JP_limits}, we expect our results to be broadly applicable
to the detection of UHE CR to which some sensitivity is expected, for which the
comparable experiment is Auger.

\section{Instantaneous Sensitivity of the Lunar Cherenkov Technique}
\label{instant}

\begin{figure*}
\centering
\includegraphics[height=5cm, clip=true]{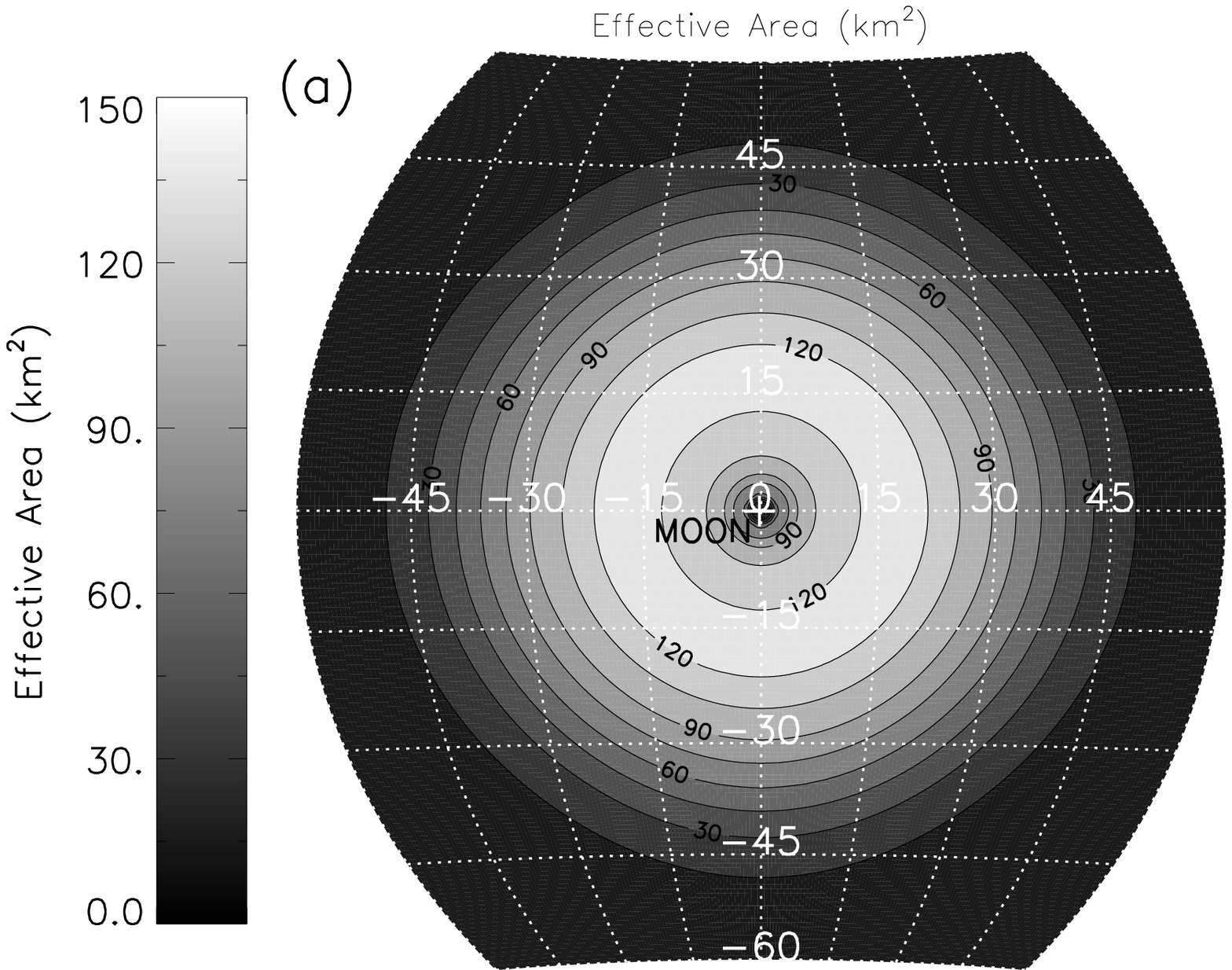} \includegraphics[height=5cm, clip=true]{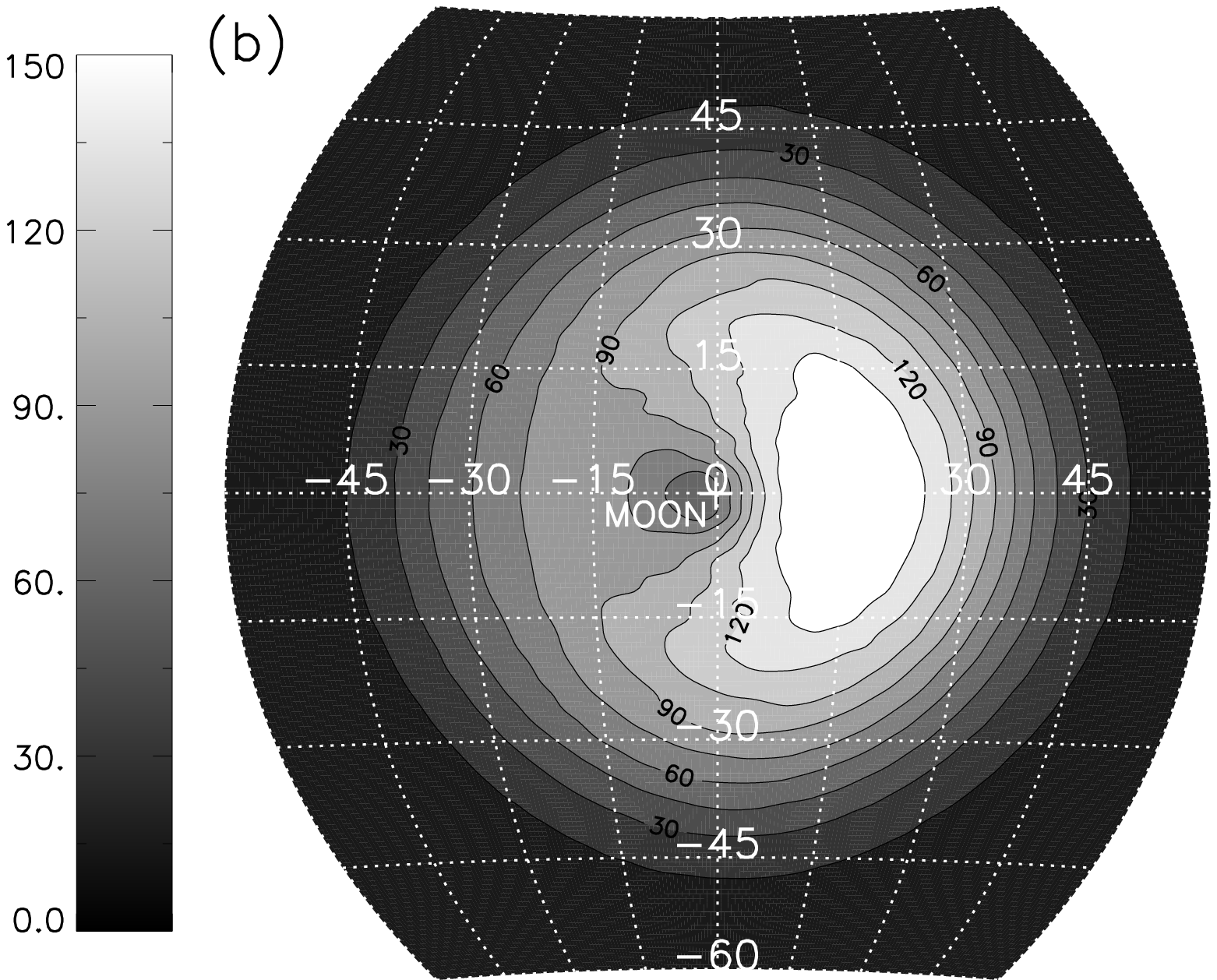}
\caption{The effective area $a_{\rm eff}$ (defined in text) of our (inferior 2007) experiment with ATCA
to $10^{22}$~eV neutrinos in both (a): centre-pointing mode, and (b): limb-pointing mode, with the beam pointing
$0.5^{\circ}$ to the right of the Moon.}
\label{instantaneous}
\end{figure*}

The most common measure of the sensitivity of Lunar Cherenkov experiments is the effective aperture
$A_{\rm eff}$ (km$^2$-sr) as a function of primary particle energy. Whether
explicitly or otherwise, this function is effectively the integral
of the (particle arrival-direction dependent) detection probability $p$ multiplied
by the effective experimental collecting area $a (\hat{\Omega}$ over the whole sky, as expressed below:
\begin{eqnarray}
A_{\rm eff}(E_{\nu}) & = & \int_{\rm sky} d \Omega \, p(E_{\nu},\hat{\Omega}) \, a (\hat{\Omega})
\end{eqnarray}
where we use $\hat{\Omega}$ to express a position on the sky in some
appropriate coordinate system.
In the case of Lunar Cherenkov experiments and accompanying simulations, the
`collecting area' ${\rm a}(\hat{\Omega})$ is best defined as the Lunar cross-section
$\pi R_m^2$ (a constant in the approximation of Lunar sphericity), and $\hat{\Omega}$
defined relative to the centre of the Moon to eliminate time-dependencies from
the Moon's motion. We define the effective area $a_{\rm eff}(\hat{\Omega}$ as follows:
\begin{eqnarray}
a_{\rm eff}(\hat{\Omega}, E_{\nu}) & = & \pi R_m^2 \, p(E_{\nu},\hat{\Omega})
\end{eqnarray}
This provides a useful measure of the sensitivity to a directionally-dependent flux.
In Fig.\ \ref{instantaneous}, we plot the calculated effective area of our $2007$
observations with ATCA (see our contribution \cite{LUNASKA_experimental})
in both centre-pointing and limb-pointing modes, using the simulation we
developed in \cite{JP_limits}.

Treating the combined antenna-Moon system as our detector, the plots in
Fig.\ \ref{instantaneous} give the effective `beam-pattern'. Though large
compared with the $\sim0.5^{\circ}$ beam width of antenna itself, nonetheless
the coverage is small when compared to the entire $4 \pi$~sr of sky. At any
one instant, the experiment will be sensitive to particles arriving only
from a small range of directions.

A useful measure of the importance of directional-dependence
in experimental sensitivity is what we define as the `directionality' ${\mathcal D}$,
the ratio of the peak effective area $a_{\rm max}$ over the mean value $\bar{a}$,
which can be related to the effective aperture as follows:
\begin{equation}
{\mathcal D} \, = \, a_{\rm max} / \bar{a} \, = \, 4 \pi \, a_{\rm max} / A_{\rm eff}
\end{equation}
For our experiment, we find values for ${\mathcal D}$ at $10^{22}$~eV of $10$ and $13$ in centre-pointing
and limb-pointing modes respectively. That ${\mathcal D}$ is higher in limb-pointing mode should
come as no surprise, since this mode sacrifices sensitivity to the majority
of events in return for increased sensitivity to a minority. Also,
as primary particle energy increases above the experimental threshold,
events with a greater range of non-optimal interactions geometries
become detectable, decreasing ${\mathcal D}$. Simulations of low-frequency experiments
indicate a reduced ${\mathcal D}$, since here the Cherenkov cone (and therefore the
acceptance) is broader at lower frequencies.

The directionality gives the possible gain in sensitivity to a point-like UHE
particle source over that to an isotropic flux (or alternatively over a blind observation).
The high values for ${\mathcal D}$ for high-frequency experiments suggest potential gains
of an order of magnitude, and thus the importance
of this type of analysis and of optimising observation times and beam-pointing positions.

\section{Current Limits on an UHE $\nu$ Flux}
\label{limits}

\begin{figure*}
\centering
\includegraphics[width=15cm, clip=true]{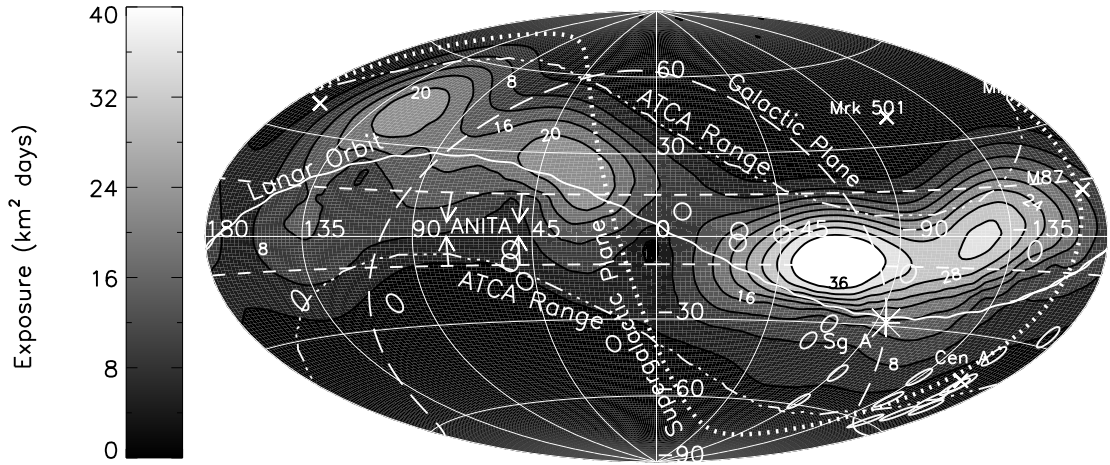}
\caption{
Combined exposure ($4,8,$\dots,$36$~km$^2$-day contours) of the GLUE and Parkes experiments
to $10^{22}$~eV neutrinos as a function of celestial coordinates. Also shown
is the $-10^{\circ} < \delta < +15^{\circ}$ declination range for ANITA \cite{ANITA}.
}
\label{exposure}
\end{figure*}

Since the instantaneous aperture of previous Lunar Cherenkov experiments
covers a small fraction of the sky, and observations have tended to be sporadic, the
limits set by these experiments are expected to be highly anisotropic.
Sufficiently accurate observation times
could only be obtained for GLUE from \cite{GLUE}, and the Parkes
experiment (our recent observations with ATCA are discussed in another
contribution \cite{LUNASKA_experimental}). Integrating the calculated
instantaneous sensitivity (with the appropriate
orientation and pointing position) over this observation time produces
the exposure map of Fig.\ \ref{exposure}, shown for $10^{22}$~eV neutrinos.

The greater contribution comes from the Goldstone experiment, where
the effect of pointing at the Northern limb of the Moon is obvious. The peak
sensitivity however lies in the declination range accessible to ANITA/ANITA-lite,
which has set the strongest limits at this energy. Other experiments with significant
limits to $10^{22}$~eV neutrinos are RICE and the Kalyazin experiment.
Since RICE is mostly sensitive to Southern latitudes,
it will not compete with Lunar Cherenkov contributions in these regions.
To UHE neutrinos arriving from the declination of Sgr A* ($-29^{\circ}$), RICE
has a maximum exposure of approximately $340$~km$^2$-days, and $260$~km$^2$-days to Cen A
at $-43^{\circ}$  \cite{RICE}. For Kalyazin,
the exposure is expected to be similar to that for GLUE given the similarity of
the experiments, though probably more spread over the lunar cycle.

Evidently from Fig.\ \ref{exposure}, significant regions of the sky
have been relatively unprobed by UHE neutrino detection experiments,
particularly in the Northern hemisphere.
Our recent observations with ATCA \cite{LUNASKA_experimental}
have begun targeting the region outside ANITA's in the South, around Sgr A* and Cen A.
A significant contribution can be made
by lunar Cherenkov experiments via a careful choice of observation times and beam pointing
positions in cases where the isotropic sensitivity may not be competitive.

\section{Future Experimental Exposure}
\label{applications}

Future lunar Cherenkov experiments planned with ATCA, ASKAP, LOFAR, and the SKA
should constitute dedicated observing runs over the entire lunar month. However,
observations will still be limited by the constraints of the Moon's orbit, which
is inclined at $\sim5^{\circ}$ to the ecliptic with nodal precession period $18.6$
years. Therefore the potential exposure to putative UHE particle
sources will be a function of angular distance from plane of the Lunar orbit.

\begin{figure}
\centering
\includegraphics[width=3.0in]{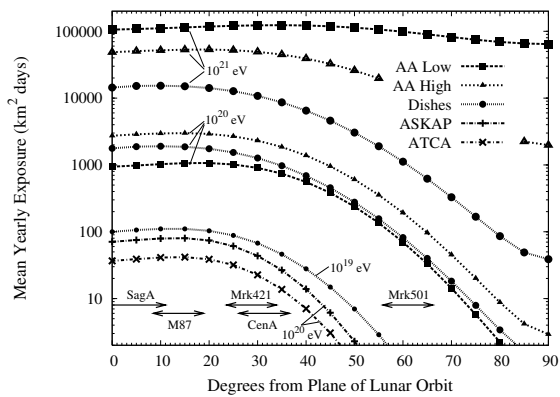}
\caption{
(Figure taken from James \& Protheroe 2008 \cite{JP_directional}) Calculated exposure for a calendar year's worth of observations for likely future experiments with a fully upgraded ATCA, ASKAP, and the SKA.
}
\label{future}
\end{figure}

In Fig.\ \ref{future} we plot our calculated exposures for a calender year's 
worth of observing for a fully optimised ATCA (not our current experiment),
ASKAP, and various SKA frequency ranges as a
function of this angular distance, and for comparative purposes include the
position of `interesting' astronomical objects, though we do not intend this
to be a complete list of potential sources (see \cite{JP_directional}).

It is particularly of interest that only for very low frequency experiments such
as our modelled $70-200$~MHz component of the SKA does the exposure become uniform,
and then only at energies at $10^{21}$~eV and above. Below this energy, all foreseeable lunar
Cherenkov experiments will have excess sensitivity within $\sim30^{\circ}$ of the lunar orbit,
or half the sky. There will thus exist an unprobed gap in the primary energy--arrival direction
parameter space near the ecliptic poles below $10^{21}$~eV, which will require
alternative detection methods to fill.

\section{Conclusions}
\label{conclusions}
We have calculated the effective `beam-pattern' of the antenna-Moon system
for a range of lunar Cherenkov experiments, finding both the instantaneous
sensitivity to, and past limits on, UHE neutrinos to be highly anisotropic.
Our results show how the sensitivity of Lunar Cherenkov observations
to potential astrophysical sources of UHE particles may be maximised
by choosing appropriate observations dates and antenna-beam
pointing positions. The likely coverage of
future experiments will be broad, but will leave large regions near
the ecliptic poles unprobed to $E<10^{21}$~eV neutrinos.

\section{Acknowledgments}

This research was supported by the Australian
Research Council's Discovery Project funding scheme (project
numbers DP0559991 and DP0881006).

\end{document}